\documentclass{article}%
\usepackage{amsfonts}
\usepackage{amsmath}
\usepackage{amssymb}
\usepackage{graphicx}%
\begin{document}

\title{The dependence of the "experimental" pion-nucleon sigma term on higher partial waves}
\author{J. Stahov\\Abilene Christian University, Abilene, TX, 76999, USA\\\ \ \ \ \ \ \ \ \ \ \ \ \ \ \ \ \ \ and\\University Tuzla, 35000 Tuzla, Bosnia and Herzegovina}
\maketitle

\begin{abstract}
A dependence of the value of the $\pi N$ sigma term on higher partial waves is
discussed. Two recent predictions of a high value of sigma term are
scrutinized. It has been concluded that the main reason for obtaining high
values of the sigma term are input D waves that are not consistent with analyticity

\end{abstract}

\section{Introduction}

The value of the $\pi N$ sigma term $\Sigma$ is given in terms of the $\bar
{D}^{+}$amplitude (bar indicates that the pseudovector Born term is
subtracted) at the Chang-Dashen (CD) point $\nu=0,t=2m_{\pi}^{2}$:%
\[
\Sigma=F_{\pi}^{2}\bar{D}^{+}(\nu=0,t=2m_{\pi}^{2}),
\]
where $F_{\pi}$=92.4 MeV is the pion decay constant. For details concerning
the $\pi N$ kinematics we refer to reference\cite{ho}. Generally, there are
two kinds of methods used to calculate the $\bar{D}^{+}$amplitude at the CD
point. The first method uses dispersion relations to calculate $\bar{D}%
^{+}(0,2m_{\pi}^{2})$ directly. The second method determines the coefficients
in the subthreshold expansion of the $\bar{D}^{+}$ amplitude:%
\begin{align*}
\bar{D}^{+}(0,2m_{\pi}^{2})  &  =(\bar{d}_{00}^{+}+\bar{d}_{01}^{+}t+\bar
{d}_{02}^{+}t^{2}+..)\text{,}\\
\text{\ }\Sigma &  =(\bar{d}_{00}^{+}+\bar{d}_{01}^{+}t+\bar{d}_{02}^{+}%
t^{2}+..)F_{\pi}^{2},\\
\Sigma &  \equiv\Sigma_{D}+\Delta_{R},
\end{align*}
where $\Sigma_{D}$ denotes contributions from the first two terms, and
$\Delta_{R}$ is so called curvature term that includes contributions of
quadratic and higher terms. Obtained values for $\Sigma$ range from \ \ 60
$MeV$ to 93 $MeV.$

It is of interest to understand which partial waves give important
contributions to $\bar{D}^{+}$ in each of the above mentioned methods. It is
clear that the leading contributions come from the input S and P waves, but
earlier evaluations(see reference\cite{ho} and references therein) show that
contributions from D and higher partial waves must not be neglected.

\section{The role of the higher partial waves}

In order to demonstrate the importance of the higher partial waves \ in
determination of the $\Sigma$ term, let's briefly describe two methods that
have produced dramatically different results in the past few years. Gasser,
Leutwyler, Locher, and Sainio(\textit{GLLS})\cite{glls} proposed a method to
improve results for $\Sigma_{D}$ previously derived from the KH80\cite{kh80}
solution by taking into account newer, mutually consistent meson factory data
below pion lab. momentum \textit{k}$_{0}=185$ $MeV/c$. The method is based on
\ six forward dispersion relations for the invariant amplitudes \textit{D}%
$^{\pm}$, $B^{\pm},E^{\pm}\equiv\frac{\partial}{\partial t}D^{\pm}.$ \ D waves
and higher partial waves are needed as a part of the input below a cutoff
momentum \textit{k}$_{0}.$ \ Above \textit{k}$_{0},$ results from one of the
existing PW solutions are used. As a result, \textit{GLLS} machinery predicts
coefficients $\bar{d}_{00}^{+}$ and $\bar{d}_{01}^{+}$ in the subthreshold
expansion. The curvature term $\Delta_{R}$ was determined using another
method\cite{opetgasser}. Several further updates were made by Sainio. In
reference\cite{sainio} a value $\Sigma=62$ $MeV$ was reported. Higher partial
waves below cutoff momentum \textit{k}$_{0}$ were taken from Ka85
solution\cite{koch85}. It was pointed out that \textquotedblleft results are
rather insensitive to the choice of PW solution \textit{above }the cutoff
momentum\textquotedblright. The most recent update was given in
reference\cite{sainiochd}. Results from Sp00 solution\cite{said},
\textit{including D waves below k}$_{0},$ were used. Obtained value,
$\Sigma=93$ \textit{MeV}, is more than 50\% higher than previously reported
values. One concludes that \textit{GLLS} machinery is sensitive to the input
for D waves and higher partial waves below the cutoff momentum.%
\begin{figure}
[p]
\begin{center}
\includegraphics[
height=2.5438in,
width=3.628in
]%
{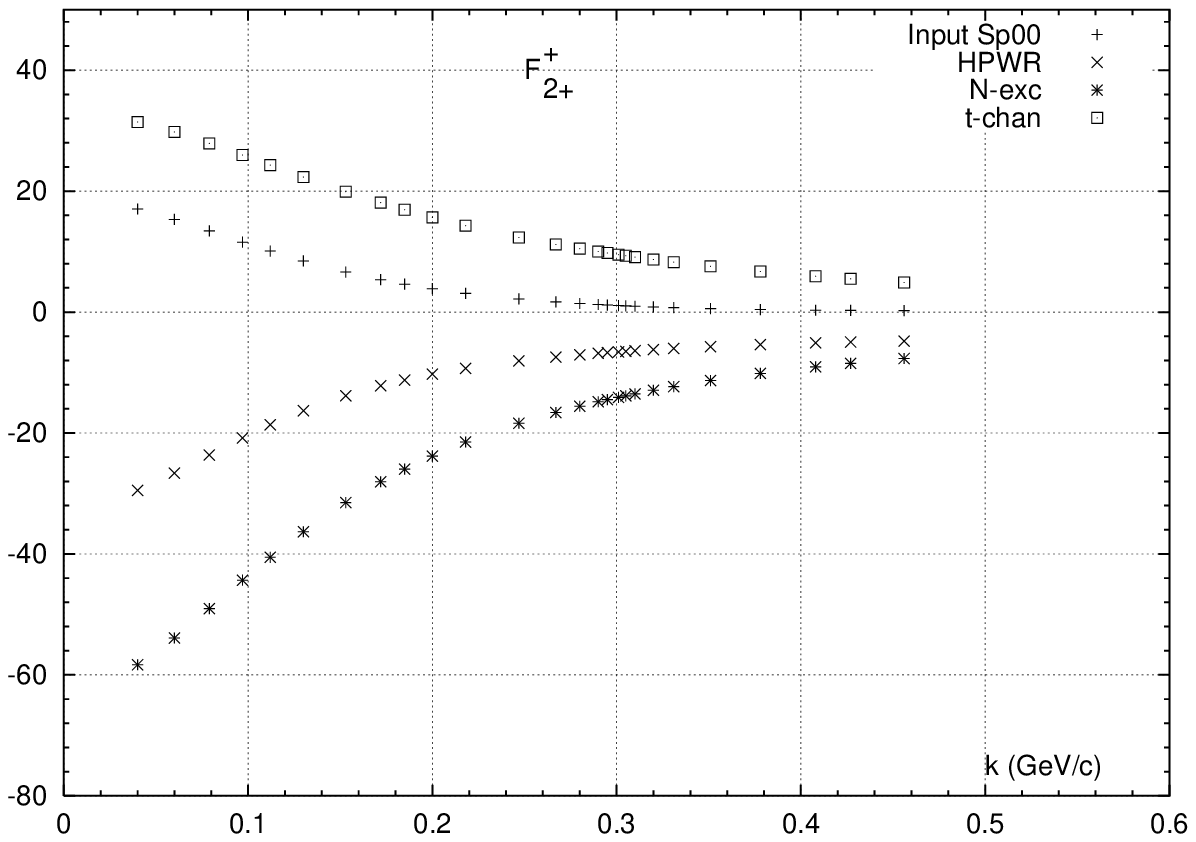}%
\caption{Comparizon of F$_{2+}^{+}$ from HPWR to input from Sp00.}%
\end{center}
\end{figure}

Starting from the fixed-\textit{t} dispersion relations for $t=2m_{\pi}^{2}$,
Olsson\cite{olsson} derived a sum rule in which the value of the $\bar{D}^{+}$
amplitude at the CD point is expressed in terms of the S, P, D and higher
partial waves threshold parameters. Using Koch's\cite{koch85} values for the D
and higher partial wave scattering lengths, Olsson obtained value
$\Sigma=(71\pm9)MeV.$ Using \textit{D-and higher partial wave scattering
lengths from VPI/GW solution Sm01\cite{said}}, Olsson and
Kaufmann\cite{olmenu} recently obtained significantly higher values ranging
from 80 $MeV$ to 88 $MeV$.

Common to both of these high evaluations of the sigma term was the use of the
higher partial waves at low energy from the latest VPI/GW solutions.

The dependence of the sigma term on the higher partial waves can be easily
seen when applying Olsson sum rule. D-wave scattering lengths from the VPI/GW
solution Sp00 lead to the sigma term value that is roughly 19 MeV higher than
value in reference\cite{olsson} where D waves from Ka84 solution have been
used. The VPI/GW group recently made \ some progress toward the lower values
of the sigma term. D waves in their solutions Sm01 and Sm02 \ are smaller but
still much higher than those from Ka84 solution. In Olsson sum rule
\ contribution of the Sm02 D waves to the $\bar{D}^{+}$ amplitude at CD point
is 0.13 MeV$^{-1}($ compared to $-0.05MeV^{-1}$ from Ka84$),$ that makes the
sigma term roughly 11 MeV higher than in reference\cite{olsson}.%

\begin{figure}
[p]
\begin{center}
\includegraphics[
height=2.5438in,
width=3.628in
]%
{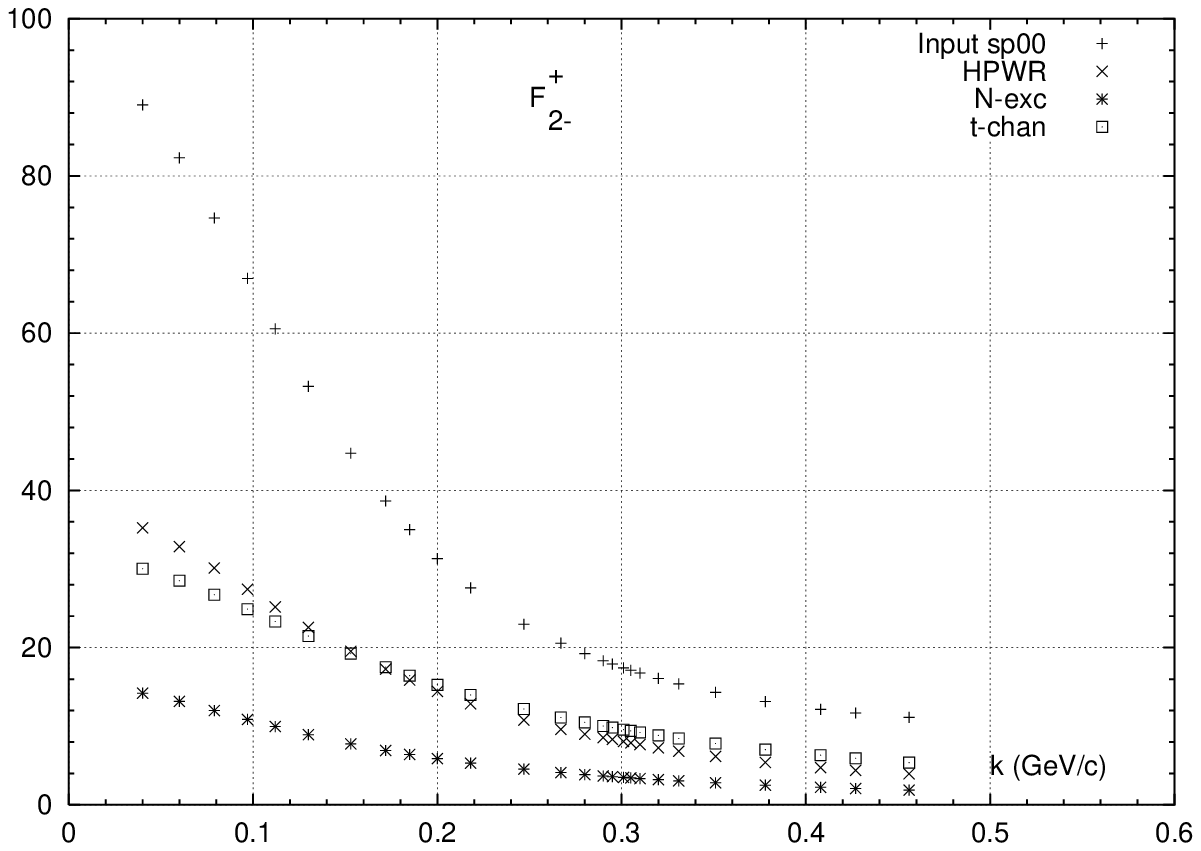}%
\caption{Comparizon of F$_{2-}^{+}$ from HPWR to input from Sp00.}%
\end{center}
\end{figure}

It is important to point out that below the \textit{GLLS} cutoff momentum
$k_{0}=185$ $MeV/c$ reliable values for D and higher partial waves can not be
obtained from experimental data only. In another words, increase of the value
of the sigma term due to contributions of D waves has no firm experimental
foundation. D waves from VPI/GW solutions should be checked. One has to start
from the first principle in $\pi N$ physics-Mandelstam analyticity and the
analytic structure of the $\pi N$ partial waves. Consistency\ of a given
partial wave with analyticity can be tested using one of the methods developed
in the past (see reference\cite{jamenu} and references therein).

\section{Hyperbolic partial wave relations as a test of higher partial waves}

In the hyperbolic partial wave relations (\textit{HPWR})\cite{steiner}, a
given $s$-channel partial wave is expressed in terms of other $s$-channel
partial waves and the $t$-channel partial waves, multiplied by corresponding
$s$-channel and $t$-channel kernels. Explicitly known kernels in the HPWR
reproduce the analytic structure of the $\pi N$ partial waves. In addition,
there is also a contribution from a nucleon exchange term that is explicitly
known as well. The method is superior compared to other methods when
predicting higher partial waves (D waves and higher) in the low energy region
($k\leq500$ $MeV/c).$ In that case, there are two leading contributions, the
nucleon exchange and the \textit{t}-channel contribution\cite{japrep}.%
\begin{figure}
[p]
\begin{center}
\includegraphics[
height=2.5438in,
width=3.628in
]%
{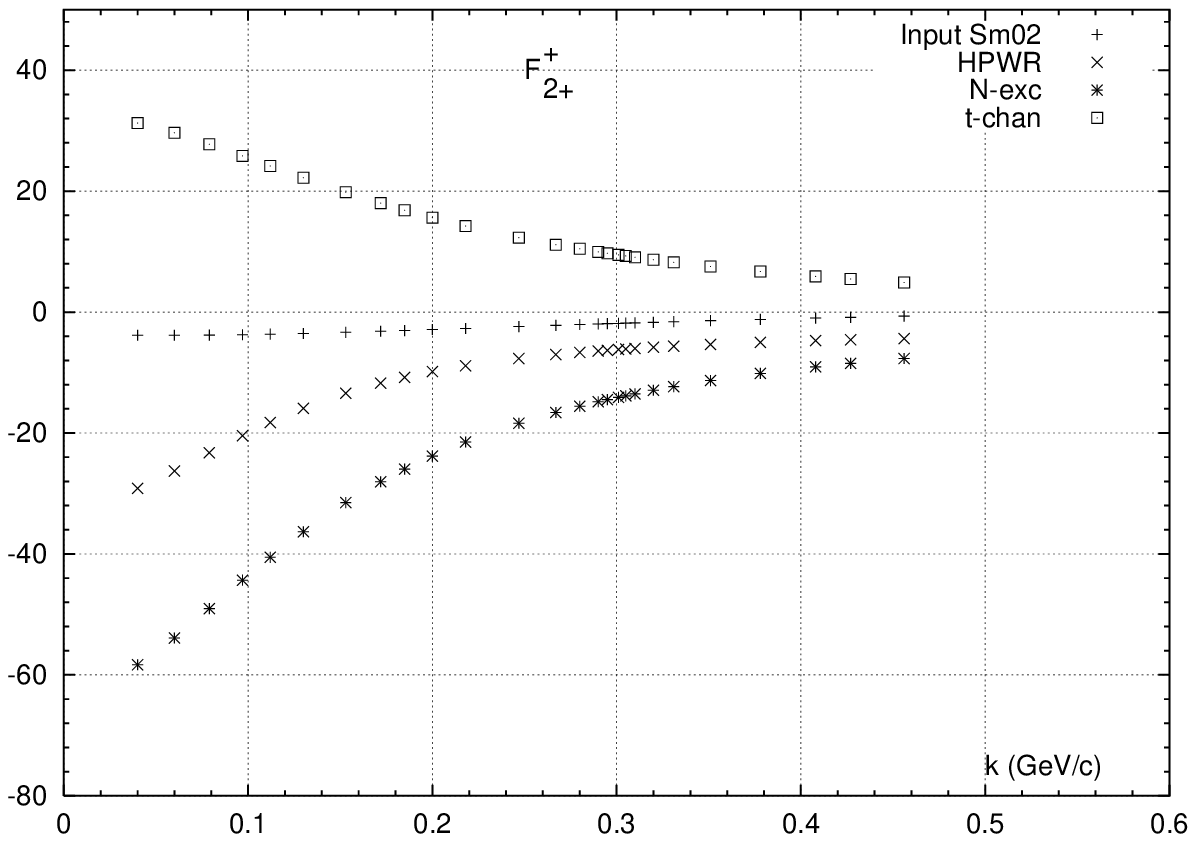}%
\caption{Comparizon of F$_{2+}^{+}$ from HPWR to input from Sm02.}%
\end{center}
\end{figure}
Due to behavior of the \textit{t}-channel kernels, contributions from higher
values of $t$ are strongly suppressed so that input available today makes it
possible to obtain reliable predictions for the higher $\pi N$ partial waves.
For example, the main contributions to the isospin even combinations of D
waves, $F_{2\pm}^{+},$ come from the kinematical region \textit{t}%
$\leq25m_{\pi}^{2}.$ \ Recent calculation\cite{jatch} shows that our input
from the \textit{t}-channel in that kinematical region is fairly well known.
Results from \textit{HPWR} for the isospin even combinations of reduced D
waves $(F_{l\pm}^{+}=T_{l\pm}^{+}/q^{l+1}$ , see ref.\cite{ho}$)$ are shown in
Fig. 1 and Fig. 2. It is evident that D waves from VPI/GW solutions Sp00 and
Sm02 are not consistent with analyticity at low energies. \ For instance, the
isospin even combination F$_{2+}^{+}$ (shown in Fig. 1) has the wrong sign. D
waves from Sm02 solution are smaller but still far from being consistent with
analyticity. D waves from the Sm02 solution are compared to HPWR calculations
in Figures 3 and 4.

\textit{ }It is hard to understand the fact that VPI/GW group\cite{pavanmenu}
constraints their P waves to the approximate Chew Low theory\cite{hamilton}%
(lowering their value of the sigma term for 6 MeV), completely ignoring at the
same time results for D waves in Ka84 that are constrained by exact PWDR.

%

\begin{figure}
[p]
\begin{center}
\includegraphics[
height=2.5438in,
width=3.628in
]%
{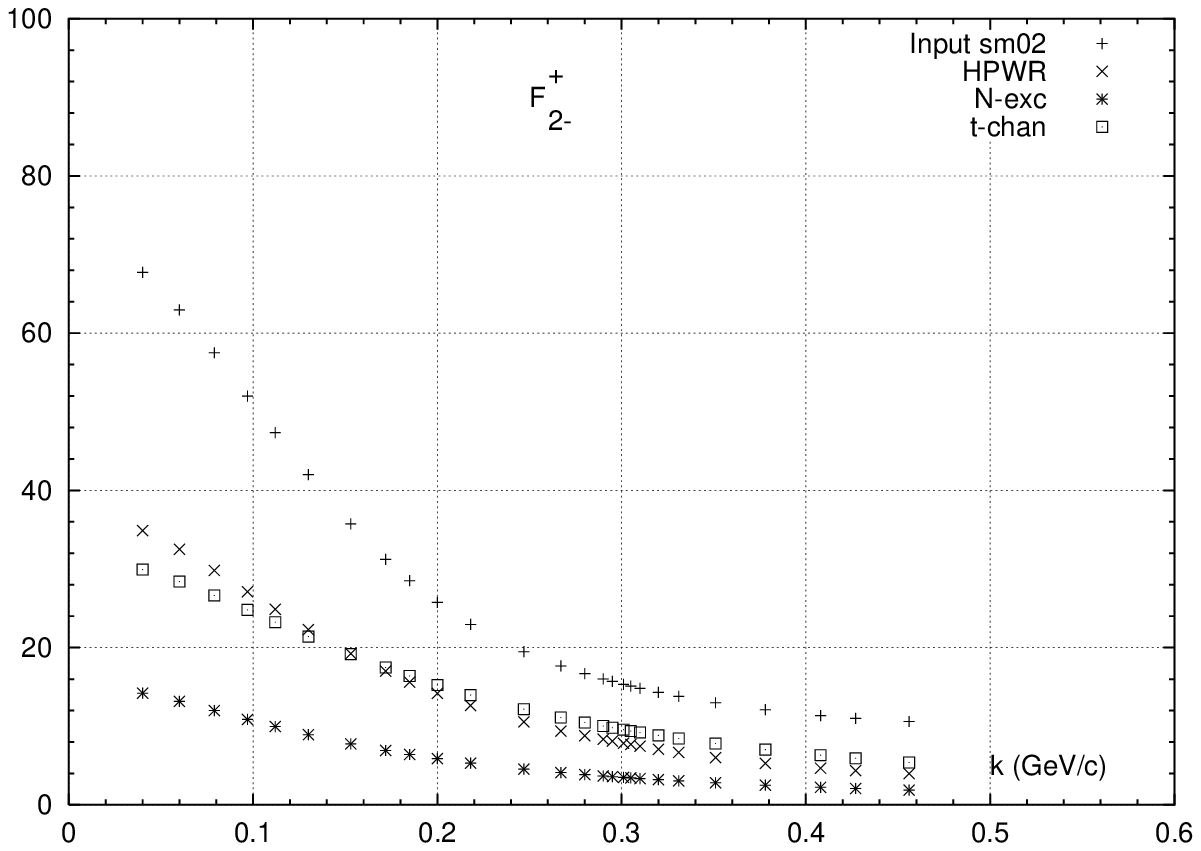}%
\caption{Comparizon of F$_{2-}^{+}$ from HPWR to input from Sm02.}%
\end{center}
\end{figure}
\newpage

\section{Conclusions}

D waves in the VPI/GW Sp00 and Sm02 partial wave solutions are not consistent
with analyticity in the low energy region and are evidently wrong. Methods
sensitive to the input D waves produce high values of the pion nucleon sigma
term because of the low energy D waves from VPI/GW solutions. A high
\textquotedblleft experimental value\textquotedblright\ of the sigma term
could be accepted\ as reliable only if partial waves from the\ input partial
wave solution are consistent with analyticity.

\bigskip

\textit{I wish to thank Prof. G. H\"{o}hler for his continuous support and
valuable discussions. This work was supported by DOE project
DE-FG03-94ER40860}

\bigskip

\bigskip

\bigskip
\end{document}